\documentclass[letterpaper,12pt]{JHEP3}
\usepackage{amsfonts, amsmath, amssymb, amsthm, cite}
\usepackage[english]{babel}

\newcommand{\eq}{\begin{equation}}
\newcommand{\feq}{\end{equation}}
\newcommand{\eqn}{\begin{eqnarray}}
\newcommand{\feqn}{\end{eqnarray}}

\title{Nonextremal black holes in gauged supergravity and the real formulation of special geometry II}

\author{Dietmar Klemm$^{ab}$ and Owen Vaughan$^{c}$ \\
$^a$ Dipartimento di Fisica, Universit\`a di Milano, \\
\hspace*{0.15cm} Via Celoria 16, 20133 Milano, Italy. \\
$^b$ INFN, Sezione di Milano, Via Celoria 16, 20133 Milano, Italy. \\
$^c$ Department of Mathematical Sciences, \\
\hspace*{0.15cm} University of Liverpool, Peach Street, Liverpool L69 7ZL, UK.
}
\preprint{IFUM-1002-FT\\ LTH 962}

\abstract{In arXiv:1207.2679 a new prescription for finding nonextremal black hole solutions
to ${\cal N}=2$, $D=4$ Fayet-Iliopoulos gauged supergravity was presented,
and explicit solutions of various models containing one vector multiplet were constructed.
Here we use the same method to find new nonextremal black holes to more complicated models.
We also provide a general recipe to construct non-BPS extremal solutions for an
arbitrary prepotential, as long as an axion-free condition holds. These follow from a set of first-order
conditions, and are related to the corresponding
supersymmetric black holes by a multiplication of the charge vector with a constant field rotation
matrix $S$. The fake superpotential driving this first-order flow
is nothing else than Hamilton's characteristic function in a Hamilton-Jacobi formalism, and coincides
in the supersymmetric case (when $S$ is plus or minus the identity) with the superpotential proposed by
Dall'Agata and Gnecchi in arXiv:1012.3756. For the nonextremal black holes that asymptote to
(magnetic) AdS, we compute both the mass coming from holographic renormalization and the one
appearing in the superalgebra. The latter correctly vanishes in the BPS case, but also for certain values of
the parameters that do not correspond to any known supersymmetric solution of ${\cal N}=2$ gauged
supergravity. We finally show that the product of all horizon areas depends only on the charges and
the asymptotic value of the cosmological constant.
}


\begin{document}

\section{Introduction}

Black holes in anti-de~Sitter (AdS) spaces play an important role in the AdS/CFT correspondence.
In the BPS case, for instance, their conformal boundaries can provide curved backgrounds
on which supersymmetric field theories can be defined \cite{Festuccia:2011ws,Cassani:2012ri}. 
On the other hand, nonextremal black holes are instrumental in recent AdS/CFT applications to
condensed matter physics, since they are dual to certain condensed matter systems at finite
temperature. A basic ingredient of realistic condensed matter systems is the presence of a finite density
of charge carriers, which implies the necessity of a bulk $\text{U}(1)$ gauge field. A further step in modeling
strongly coupled holographic systems is to include the leading relevant (scalar) operator in the
dynamics. This is generically uncharged, and is dual to a neutral scalar field in the bulk.
One is thus naturally led to consider nonextremal charged black holes in gauged supergravity
with running scalars.

AdS can be viewed as a deformation of Minkowski spacetime, and one would like to understand how
much of the features of asymptotically flat black holes carry over to their cousins in AdS, or,
more generally, how certain recipes to construct such solutions get modified when one turns on
a cosmological constant or a potential for the moduli. For example, it is known how to go from
vanishing to nonzero temperature for asymptotically flat black holes and black branes, namely
by introducing a so-called blackening function in certain components of the metric. Although
something similar seems to work in special cases \cite{Behrndt:1998jd}, the AdS analogue of this
procedure is unknown in general.

Recently there has also been an increasing interest in the attractor mechanism in presence
of a scalar potential \cite{Bellucci:2008cb,Cacciatori:2009iz,Dall'Agata:2010gj,Kachru:2011ps,Inbasekar:2012sh}. Unless the latter has flat directions, the values of the moduli at
infinity will be completely fixed, and can thus not be continuously varied. It is therefore clear
that the attractor mechanism in this case will be qualitatively different from the one for asymptotically
flat black holes. In fact, new properties emerge in gauged supergravity, for instance possible nontrivial
moduli spaces for the BPS attractor flow \cite{Cacciatori:2009iz}. This is in contrast to ungauged
supergravity, where there are no flat directions in the black hole potential for BPS
flows \cite{Ferrara:1997tw} (at least as long as the metric of the scalar manifold is strictly positive definite),
but there is a nontrivial moduli space for non-BPS flows \cite{Ferrara:2007tu,Bellucci:2008sv}.

A further point that motivates studying AdS black holes is the existence of solitons, i.e., there are
solutions that have no smooth limit when the coupling constant goes to zero, like Romans'
cosmic monopole \cite{Romans:1991nq} and its generalizations \cite{Caldarelli:1998hg,Cacciatori:2009iz}.

For these reasons, it would be desirable to dispose of a more complete picture of possible asymptotically AdS
black holes, at least for supergravity models that can be consistently embedded in string theory. One such
model is four-dimensional ${\cal N}=2$ gauged supergravity with prepotential $F=-2i(X^0X^1X^2X^3)^{1/2}$,
that will be considered below. Known black hole solutions to this theory, which contains three
vector multiplets, include the static four-charge black holes of \cite{Duff:1999gh} and the rotating ones
of \cite{Chong:2004na}, that have two pairwise equal electric charges. However, the most general rotating
solution with nut-charge, four electric and four magnetic charges, containing also the BPS black holes of
\cite{Cacciatori:2009iz,Klemm:2011xw,Colleoni:2012jq}, has not been constructed so
far\footnote{The supersymmetric solutions of \cite{Colleoni:2012jq}, which were found by using the
general recipe provided in \cite{Cacciatori:2008ek}, contain two pairwise equal
electric and magnetic charges as well as nut charge. These are not all independent, but are given in
terms of three free parameters.}.

In \cite{Klemm:2012yg} a new prescription for finding nonextremal black hole solutions
to ${\cal N}=2$, $D=4$ Fayet-Iliopoulos gauged supergravity was presented,
and explicit solutions of various models containing one vector multiplet were constructed.
Here we shall go one step further, and use the same method to find new nonextremal black holes to
more complicated models.

The remainder of this paper is organized as follows: In section \ref{review} we review the recipe
of \cite{Klemm:2012yg}, which made essential use of the formalism developed in \cite{Mohaupt:2011aa},
based on dimensional reduction and the real formulation of special geometry. In
section \ref{sqrtX0X1X2X3-model}, the $F=-2i(X^0X^1X^2X^3)^{1/2}$ model is considered, and new
nonextremal black holes are constructed. They carry four magnetic charges, and contain
both the solutions of \cite{Duff:1999gh} and the supersymmetric ones found in \cite{Cacciatori:2009iz}.
We compute their entropy, temperature, the mass coming from holographic renormalization and the one
appearing in the superalgebra. The latter correctly vanishes in the BPS case, but quite surprisingly
also for certain values of the parameters that do not correspond to any known supersymmetric solution
of ${\cal N}=2$ gauged supergravity. It is moreover shown that the product of all horizon areas depends
only on the charges and the asymptotic value of the cosmological constant, confirming the universality
of the area-product formula \cite{Cvetic:2010mn}. In the following section, we consider the model
with prepotential $F=-X^1X^2X^3/X^0$. In the ungauged case, this is related to the one of section
\ref{sqrtX0X1X2X3-model} by a symplectic transformation, but since the gauging breaks symplectic
covariance, this will lead to new physics. In fact, unlike the solutions of \ref{sqrtX0X1X2X3-model},
those that we shall construct in \ref{-X1X2X3/X0-model} do not asymptote to AdS.

For both models, we find that an extremal subclass of these solutions\footnote{Actually, for the
$F=-X^1X^2X^3/X^0$ model, we only find extremal solutions.}, which is related to the corresponding
supersymmetric black holes by a simple individual sign-flipping of the charges,
follows from a set of first-order conditions. The fake superpotential appearing in these first-order equations
is nothing else than Hamilton's characteristic function in a Hamilton-Jacobi formalism.
It is shown in section \ref{W-arbitr-prepot} that this can be generalized to any prepotential, as long as
a zero-axion condition holds. This provides a general recipe that allows to construct non-BPS
extremal black holes in ${\cal N}=2$ FI-gauged supergravity. We obtain the fake superpotential $W$,
and show that in the supersymmetric case (when the so-called field rotation matrix is plus or minus
the identity), $W$ coincides with the true superpotential proposed by Dall'Agata and Gnecchi
in \cite{Dall'Agata:2010gj}.

We conclude in \ref{conclusions} with an outlook and some final remarks.

\section{Review of method}
\label{review}

In this section we review the prescription for finding nonextremal solutions presented in \cite{Klemm:2012yg}, which is based on dimensional reduction over the timelike dimension and the formulation of projective special
K\"ahler geometry in terms of real rather than complex coordinates. The final ingredient is a coordinate redefinition that allows the moduli fields to absorb the Kaluza-Klein scalar (a.k.a.\ the metric warp factor) in order to lift a hypersurface constraint. For more information on this procedure we refer the reader to \cite{Mohaupt:2011aa}, in which this approach was developed for four-dimensional ${\cal N} = 2$ (ungauged) supergravity for the first time, and new rotating and static solutions to the ungauged theory were discussed. 

This approach was originally developed in \cite{Mohaupt:2009iq} for five-dimensional supergravity. In the same paper new multi-centered extremal solutions were found, which have subsequently been generalised to non-extremal solutions for certain classes of models \cite{Mohaupt:2010fk,Mohaupt:2012tu}. When restricted to to static and spherically symmetric backgrounds the equations of motion also take a similar in form to those recently presented in \cite{Meessen:2011aa}.

\subsection{Equations of motion for generic models}

	Our starting point is the Lagrangian given by expression (2.7) of \cite{Klemm:2012yg}:
	\begin{align}
			{\cal L}_3 \;\sim\;&  - \tfrac{1}{2} R_3 - \tilde{H}_{ab} \left(\partial_\mu q^a \partial^\mu q^b - \partial_\mu \hat{q}^a \partial^\mu \hat{q}^b \right) + \frac{1}{2H} V \;, \label{eq:3dLag_static} 
 	\end{align}
which is a three-dimensional effective Lagrangian for a theory of Fayet-Iliopoulos gauged supergravity coupled to $n$ vector multiplets on static backgrounds. It is understood that the four-dimensional spacetime metric has been decomposed in the standard manner adapted to dimensional reduction,
\[
	ds^2_4 = -e^\phi (dt^2 + V_\mu dx^\mu) + e^{-\phi} ds^2_3 \;.
\]
We refer to $e^\phi$ as the Kaluza-Klein scalar and $V_\mu$ as the Kaluza-Klein vector. The latter vanishes for static backgrounds when appropriate choices of coordinates are made, which we will assume from now on.
The scalar fields $q^a = (x^I, y_I)^T$ appearing in (\ref{eq:3dLag_static}) represent the combined degrees of freedom of both the complex scalar fields $X^I,\bar{X}^I$ and the KK-scalar $e^{\phi}$,
	\begin{equation} \label{xy}
	\begin{aligned}
		x^I &:= \text{Re}\, Y^I = e^{\phi/2} \text{Re}\, X^I  \;,  \\
		y_I &:= \text{Re}\, F_I(Y) = e^{\phi/2} \text{Re}\, F_I(X)  \;.
	\end{aligned}
	\end{equation}
	Here we have implicitly defined $Y^I,F_I$ as rescaled versions of $X^I,F_I$.
	The scalar fields $\hat{q}^a = ( \frac{1}{2} \zeta^I, \frac{1}{2} \tilde{\zeta}_I )^T$ descend from the degrees of freedom of the gauge fields,
	\begin{equation} \label{eq:field_strengths}
	\begin{aligned}
		\partial_\mu \zeta^I &:= F_{\mu 0}^{I}  \;, \\
		\partial_\mu \tilde{\zeta}_I &:= G_{I|\mu 0} \;.
	\end{aligned}
	\end{equation}
The scalar couplings are completely determined by the Hesse potential $H(x,y)$, which plays the role of the holomorphic prepotential $F(X)$ when using real coordinates. The two potentials are related by a Legendre transformation that replaces $\text{Im}(X^I)$ with $\text{Re}(F_I)$ as an independent field \cite{Cortes:2001}. The Hesse potential also happens to be proportional to the K\"ahler potential\footnote{The exact relation is $-2H = e^{\cal -K}$.}. The coupling matrix that appears in (\ref{eq:3dLag_static}) is given in terms of the Hesse potential by
\[
	\tilde{H}_{ab} = \frac{\partial^2 \tilde{H}}{\partial q^a \partial q^b} \;, 
\]
where $\tilde{H} =  -\frac12 \log -2 H$.		
The function $V$ appearing in the Lagrangian is nothing other than the Fayet-Iliopoulos potential, and the whole term $\frac{1}{H}V$ can be written in terms of the Hesse potential as
\begin{equation}
	\frac{1}{H}V(q) = 2 g_I g_J \left( -\frac{\partial^2}{\partial y_I \partial y_J} \tilde{H} + \frac{2}{H^2} x^I x^J + 4 \frac{\partial \tilde{H}}{\partial y_I} \frac{\partial \tilde{H}}{\partial y_J}\right) \;.
	\label{eq:potential_H}
\end{equation}
This simple expression for this potential is another major advantage of using real coordinates and their associated Hesse potential.
The last comment we need to make about the Lagrangian is that the equations of motion must be supplemented by the additional constraints
	\begin{equation}
		q^a \Omega_{ab} \partial_\mu q^b = q^a \Omega_{ab} \partial_\mu \hat{q}^b = 0 \;,
		\label{additional-ansatz}
	\end{equation}	
	and the integrability condition
		\begin{equation}
		\partial_\mu \tilde{\phi} + 2\hat{q}^a \Omega_{ab} \partial_\mu \hat{q}^b = 0 \;. \label{static}
	\end{equation}
The integrability condition corresponds to the requirement that the field strength of the Kaluza-Klein vector vanishes, i.e.\ solutions are static. For black hole solutions to the ungauged theory the constraints (\ref{additional-ansatz}) correspond to a vanishing NUT charge \cite{Mohaupt:2011aa}.

We now perform the variation of the Lagrangian (\ref{eq:3dLag_static}) with respect to the fields $q^a,\hat{q}^a$ and $g_{3\mu\nu}$, and obtain the respective equations of motion
	\begin{align}
			& \nabla^\mu \left[\tilde{H}_{ab}\partial_\mu q^b \right] -  \tfrac{1}{2}\partial_a \tilde{H}_{bc} \left(\partial_\mu q^b \partial^\mu q^c - \partial_\mu \hat{q}^b \partial^\mu \hat{q}^c \right)  + \partial_a \left( \frac{1}{4H} V(q) \right) = 0 \;, \label{eq:eom1}\\
			&\nabla^\mu \left[\tilde{H}_{ab}\partial_\mu \hat{q}^b \right]  = 0 \;, \label{eq:eom2}\\
			&\tilde{H}_{ab} \left(\partial_\mu q^a \partial_\nu q^b - \partial_\mu \hat{q}^a \partial_\nu \hat{q}^b \right) - \frac{1}{2H} g_{\mu \nu} V(q) = -\tfrac{1}{2}R_{3 \mu \nu}  \;.\label{eq:eom3}
	\end{align}
	We solve equation (\ref{eq:eom2}) immediately by setting
	\begin{equation}
		\tilde{H}_{ab}\partial_\mu \hat{q}^b = \partial_\mu {\cal H}_a \;, \label{eg:qhat}
	\end{equation}
	where ${\cal H}_a$ are harmonic functions. Using the function $\tilde{H}$ one can define a natural set of dual coordinates which simplify the remaining equations of motion. These are defined by $q_a := \partial_a \tilde{H}$, and through the chain rule satisfy
	\begin{equation}
		\partial_\mu q_a = \tilde{H}_{ab} \partial_\mu q^b \;.
	\end{equation}
We can relate the dual coordinates to the imaginary parts of $Y^I, F_I$ using the expressions \cite{Mohaupt:2011aa}
	\[
		q_a = \frac{1}{H} \left( \begin{array}{c} -v_I \\ u^I \end{array} \right) = \frac{1}{H} \left( \begin{array}{c} -\text{Im}(F_I) \\ \text{Im}(Y^I) \end{array} \right) \;.
	\]
The remaining equations of motion can be written in terms of the dual coordinates and harmonic functions as
	\begin{align}
			\Delta q_a  +  \tfrac{1}{2}\partial_a \tilde{H}^{bc} \Big(\partial_\mu q_b \partial^\mu q_c - \partial_\mu {\cal H}_b \partial^\mu {\cal H}_c \Big)  + \partial_a \left( \frac{1}{4H} V(q) \right) &= 0 \;, \label{eq:eom_dual1}\\
			\tilde{H}^{ab} \Big(\partial_\mu q_a \partial_\nu q_b - \partial_\mu {\cal H}_a \partial_\nu {\cal H}_b \Big) - \frac{1}{2H} g_{\mu \nu} V(q) &= -\tfrac{1}{2}R_{3 \mu \nu}  \;.\label{eq:eom_dual2}
	\end{align}

\subsection{Metric ansatz}
\label{sec-metr-ans}

	Following \cite{Cacciatori:2009iz} we now make the following ansatz for the 3d metric:
	\begin{equation}
	ds_3^2 = dz^2 + e^{2\Phi(z,w,\bar w)} dw d\bar w\,,
	\label{eq:g}
	\end{equation}
	where $\Phi$ is separable,
	\eq
	\Phi(z,w,\bar w) = \psi(z) + \gamma(w,\bar w)\,, \label{Phi-separable}
	\feq
	and $\gamma$ satisfies the Liouville equation
	\eq
	\Delta_{(2)}\gamma \equiv 4\partial_w\partial_{\bar w}\gamma = -\kappa e^{2\gamma}\,, \label{liouville}
	\feq
	with $\kappa$ a constant. \eqref{liouville} means that the two-metric $e^{2\gamma}dwd\bar w$ has
	constant curvature. As a solution of \eqref{liouville} we shall take
	\eq
	e^{2\gamma} = \left[1 + \frac{\kappa}4w\bar w\right]^{-2}\,.
	\feq
	We also assume that the fields $q^a$ and $\hat{q}^a$ only depend on $z$,
	\[
	q^a = q^a(z) \;, \hspace{3em} \hat{q}^a = \hat{q}^a(z)\,.
	\]


\section{The $F(X) = -2i(X^0 X^1 X^2 X^3)^{1/2}$ model}
\label{sqrtX0X1X2X3-model}

	The Hesse potential for this model is given by (\ref{eq:STUHesse}):
	\begin{align*}
			H &= - 2 \Big( -(y_0x^0 - y_1x^1 - y_2 x^2 - y_3 x^3)^2 + 4y_1 x^1 y_2 x^2 + 4y_1 x^1 y_3 x^3 \\
				&\hspace{12em} + 4 y_2 x^2 y_3 x^3	+ 4 y_0 y_1 y_2 y_3 + 4 x^0 x^1 x^2 x^3 \Big)^{1/2} \;. 
	\end{align*}	

\subsection{Equations of motion}

	We will consider configurations of the form
	\begin{equation}
		x^0 = x^1 = x^2 = x^3 = 0 \qquad \Rightarrow \qquad v_0 = v_1 = v_2 = v_3 = 0 \;. \label{eq:restriction1}
	\end{equation}
	We call such configurations axion-free since they are related to axion free configurations of the $F = -X^1 X^2 X^3 / X^0$ model by a symplectic transformation. 
	The matrix $\tilde{H}^{ab}$ is given by
		\begin{equation}
		\tilde{H}^{ab} = \left( \begin{array}{cccccccc} 
		*	& * & * & * & 0 & 0 & 0 & 0 \\
		*	& * & * & * & 0 & 0 & 0 & 0 \\
		*	& * & * & * & 0 & 0 & 0 & 0 \\
		*	& * & * & * & 0 & 0 & 0 & 0 \\
		0 & 0 & 0 & 0 & 4{y_0}^2 & 0 & 0 & 0 \\
		0 & 0 & 0 & 0 & 0 & 4{y_1}^2 & 0 & 0 \\		
		0 & 0 & 0 & 0 & 0 & 0 & 4{y_2}^2 & 0 \\		
		0 & 0 & 0 & 0 & 0 & 0 & 0 & 4{y_3}^2		
		\end{array} \right) \;. \label{eq:Htilde1}
	\end{equation}
	Components denoted by $*$ represent possible non-zero entries. These can easily be determined but will not play a role in the discussion, as we shall see shortly.
	The equations of motion greatly simplify if we impose that in addition half of the harmonic functions are constant,
	\[
		\partial_\mu {\cal H}_0 = \partial_\mu {\cal H}_1 = \partial_\mu {\cal H}_2 = \partial_\mu {\cal H}_3 = 0 \;. 
	\]
	From the above decomposition of $\tilde{H}^{ab}$ and (\ref{eg:qhat}) we see that this condition is equivalent to switching off the electric charges. Along with the condition (\ref{eq:restriction1}) this implies that the upper-left block appearing in $\tilde{H}^{ab}$ completely decouples from the equations of motion, and are therefore no longer of any relevance.
	The non-vanishing dual coordinates read
	\begin{align}
		q_4 &= -\frac{1}{4y_0} \;, & q_6 &= -\frac{1}{4y_2} \;, \notag \\
		q_5 &= -\frac{1}{4y_1} \;, & q_7 &= -\frac{1}{4y_3} \;. \label{eq:coords_and_duals1}
	\end{align}
	The second order equations of motion \eqref{eq:eom_dual1} for the $q_a$ are given by
	\begin{align}
		&\Delta {q}_4 - \frac{\left[ (\partial_z q_4)^2 - (\partial_z {\cal H}_4)^2 \right]}{q_4} \notag \\
		&\hspace{6em} -16g_0^2 q_4^2 + 16g_0 q_4^2 \left(g_0 q_4 + g_1 q_5 + g_2 q_6 + g_3 q_7 \right) = 0 \;, \label{eq:eom1_q4} \\
		&\Delta {q}_5 - \frac{\left[ (\partial_z q_5)^2 - (\partial_z {\cal H}_5)^2 \right]}{q_5} \notag \\
		&\hspace{6em} -16g_1^2 q_5^2 + 16g_1 q_5^2 \left(g_0 q_4 + g_1 q_5 + g_2 q_6 + g_3 q_7 \right) = 0 \;, \label{eq:eom1_q5} \\ 
		&\Delta {q}_6 - \frac{\left[ (\partial_z q_6)^2 - (\partial_z {\cal H}_6)^2 \right]}{q_6} \notag \\
		&\hspace{6em} -16g_2^2 q_6^2 + 16g_2 q_6^2 \left(g_0 q_4 + g_1 q_5 + g_2 q_6 + g_3 q_7 \right) = 0 \;, \label{eq:eom1_q6} \\
		&\Delta {q}_7 - \frac{\left[ (\partial_z q_7)^2 - (\partial_z {\cal H}_7)^2 \right]}{q_7} \notag \\
		&\hspace{6em} -16g_3^2 q_7^2 + 16g_3 q_7^2 \left(g_0 q_4 + g_1 q_5 + g_2 q_6 + g_3 q_7 \right) = 0 \;, \label{eq:eom1_q7}
	\end{align}
	and the Einstein equations \eqref{eq:eom_dual2} boil down to
	\begin{align}
		&\sum_{m = 4}^7 \frac{\left[(\partial_z q_m)^2 - (\partial_z {\cal H}_m)^2 \right]}{4 q_m^2}  
		+ 4g_0^2 q_4^2 + 4g_1^2 q_5^2 + 4g_2^2 q_6^2 + 4g_3^2 q_4^7 \notag \\
		&\hspace{6em} - 4(g_0 q_4 + g_1 q_5 + g_2 q_6 + g_3 q_7)^2 = -\partial_z^2 \psi - (\partial_z \psi)^2 \;, \label{eq:eom1_E1} \\
		\notag \\
		&-8g_0^2 q_4^2 -8g_1^2 q_5^2 -8g_2^2 q_6^2 -8g_3^2 q_7^2 + 8(g_0 q_4 + g_1 q_5 + g_2 q_6 + g_3 q_7)^2 \notag \\
		&\hspace{18em} = \partial^2_z \psi +2 (\partial_z \psi)^2 - {\kappa} e^{-2\psi} \;. \label{eq:eom1_E2}
	\end{align}	
	One can check that upon making the truncation
	\begin{align*}
		&q_4 \to q_2 \;, & &g_0 \to g_0 \;, \\
		&q_5 \to \tfrac13 q_3 \;, & &g_1 \to g_1 \;, \\
		&q_6 \to \tfrac13 q_3 \;, & &g_2 \to g_1 \;, \\
		&q_7 \to \tfrac13 q_3 \;, & &g_3 \to g_1 \;,
	\end{align*}
	one obtains the equations of motion for the $t^3$ model found in \cite{Klemm:2012yg}.

Note that the fields ${\cal H}_{I+4}$ ($I=0,\ldots,3$) are harmonic functions, i.e.
\[
\partial_z {\cal H}_{I+4} = e^{-2\psi} C_{I+4}\,,
\]
where the $C_{I+4}$ are constants proportional to the magnetic charges.

\subsection{Nonextremal black holes}
\label{nonextr-bh}

In order to solve \eqref{eq:eom1_q4}-\eqref{eq:eom1_E2}, we use the ansatz of \cite{Klemm:2012yg},
\eq
q_{I+4} = \frac{f_{I+4}}{e^{\psi}}\,, \qquad f_{I+4} = z\alpha_{I+4} + \beta_{I+4}\,, \label{ansatz-q}
\feq
where $\alpha_{I+4},\beta_{I+4}$ are constants, and $e^{2\psi}$ is a quartic polynomial,
\eq
e^{2\psi} = \sum_{n=0}^4 a_n z^n\,. \label{quartic-pol}
\feq
Notice that the four-dimensional geometry has two scaling symmetries, namely
\[
(t, z, w, e^{\phi}, e^{\psi}) \mapsto (t/\mu, \mu z, \mu w, e^{\phi}\mu^2, e^{\psi})\,,
\]
and
\[
(t, z, w, e^{\phi}, e^{\psi}) \mapsto (t/\mu, \mu z, w, e^{\phi}\mu^2, e^{\psi}\mu)\,.
\]
One can use the first to set $\kappa=0,\pm 1$ (corresponding to $\mathbb{R}^2$, $\text{S}^2$ and
$\text{H}^2$ respectively) and then the second (that leaves $\kappa$ invariant) to choose $a_4=1$.
Furthermore, by shifting the coordinate $z$, it is always possible to eliminate the cubic term in
\eqref{quartic-pol}. We shall thus take $a_3=0$ in what follows. After that, it is straightforward to verify
that the equations of motion \eqref{eq:eom1_q4}-\eqref{eq:eom1_E2} are satisfied if the following
relations for the coefficients hold:
\begin{eqnarray}
\alpha_{I+4} &=& \frac1{4g_I}\,, \qquad \sum_{I=0}^3g_I\beta_{I+4} = 0\,, \qquad a_2 = \kappa
- 8\sum_{I=0}^3g_I^2\beta_{I+4}^2\,, \label{alpha-beta-a2} \\
C_{I+4}^2 &=& a_2\beta_{I+4}^2 + \frac{a_0}{16g_I^2} - \frac{a_1\beta_{I+4}}{4g_I} + 16g_I^2\beta_{I+4}^4\,,
\label{C_I+4}
\end{eqnarray}
where there is no summation over $I$ in the last equation. We are thus left with a five-parameter family
of solutions, labeled for instance by $(\beta_4,\beta_5,\beta_6,a_0,a_1)$. Note that the equations
\eqref{additional-ansatz}, \eqref{static} (with $\tilde{\phi}=\text{const}$) are trivially satisfied in this case.
The dilaton $\phi$ is computed from the Hesse potential,
\[
e^{\phi} = -2H(x,y) = \frac12 e^{2\psi}(f_4f_5f_6f_7)^{-1/2}\,.
\]
Introducing coordinates $\vartheta,\varphi$ according to
\[
w = \left\{\begin{array}{r@{\quad,\quad}l} 2\tan\frac{\vartheta}2 e^{i\varphi} & \kappa = 1 \\
                                                                                    \vartheta e^{i\varphi} & \kappa = 0 \\
                                                                                    2\tanh\frac{\vartheta}2 e^{i\varphi} & \kappa = -1
                                                                                    \end{array}\right.
\]
yields for the four-dimensional metric
\begin{equation}
ds_4^2 = -\frac{e^{2\psi} dt^2}{2(f_4f_5f_6f_7)^{1/2}} + 2(f_4f_5f_6f_7)^{1/2}\left(\frac{dz^2}{e^{2\psi}} +
d\vartheta^2 + S_{\kappa}^2(\vartheta)d\varphi^2\right)\,, \label{metric-stu}
\end{equation}
where we defined
\[
S_{\kappa}(\vartheta) = \left\{\begin{array}{c@{\quad,\quad}l} \sin\vartheta & \kappa = 1 \\
                                                                                    \vartheta & \kappa = 0 \\
                                                                                    \sinh\vartheta & \kappa = -1\,.
                                                                                    \end{array}\right.
\]
Moreover, one has from \eqref{xy}
\begin{eqnarray}
X^0 &=& -\frac{i}{2\sqrt2}f_4^{3/4}(f_5f_6f_7)^{-1/4}\,, \quad X^1 = -\frac{i}{2\sqrt2}f_5^{3/4}(f_4f_6f_7)^{-1/4}\,,
\nonumber \\
X^2 &=& -\frac{i}{2\sqrt2}f_6^{3/4}(f_4f_5f_7)^{-1/4}\,, \quad X^3 = -\frac{i}{2\sqrt2}f_7^{3/4}(f_4f_5f_6)^{-1/4}\,.
\label{X-stu}
\end{eqnarray}
Finally, from (\ref{eq:field_strengths}) the gauge field strengths read
\[
F^I_{\mu t} = G_{I|tw} = G_{I|t\bar w} = 0\,, \qquad G_{I|tz} = -\frac{C_{I+4}}{2f_{I+4}^2}\,,
\]
and using the fact that
\[
{\cal N}^{-1} = i(f_4f_5f_6f_7)^{-1/2}\text{diag}(f_4^2,f_5^2,f_6^2,f_7^2)\,,
\]
we can write this as
\eq
F^I = \frac i2 C_{I+4} e^{2\gamma} dw\wedge d\bar w\,. \label{F-stu}
\feq
Observe that the expressions for the gauge field strengths are precisely the same as for the BPS
case \cite{Cacciatori:2009iz}. The solution \eqref{metric-stu} has an event horizon at the largest
root $z_{\text h}$ of $e^{2\psi}=0$. Regularity of the Euclidean section at $z=z_{\text h}$ gives the
Hawking temperature
\[
T = \left[\frac{(e^{2\psi})'}{8\pi(f_4f_5f_6f_7)^{1/2}}\right]_{z=z_{\text h}}\,,
\]
where a prime denotes a derivative w.r.t.~$z$. Note that, if we normalize the timelike Killing vector
differently in order to have the usual asymptotically AdS behaviour (cf.~the mass computation below),
$T$ picks a prefactor, $T\to 2\sqrt2(g_0g_1g_2g_3)^{1/4}T$.
For the Bekenstein-Hawking entropy one obtains
\begin{equation}
S = \frac{A_{\text h}}{4G} = \frac{(f_4f_5f_6f_7)^{1/2}|_{z_{\text h}}V}{2G}\,,
\end{equation}
where\footnote{If the horizon is noncompact, one can still define a finite entropy density $s=S/V$.}
\eq
V \equiv \int S_{\kappa}(\vartheta)d\vartheta d\varphi\,. \label{def-vol}
\feq
If $\kappa=1$ and the Dirac-type charge quantization condition\footnote{Note that \eqref{Dirac} holds
in the supersymmetric case \cite{Cacciatori:2009iz}, due to the minimal coupling of the gravitinos
to the linear combination $g_IA^I$.}
\eq
2\sum_{I=0}^3g_IC_{I+4} = \pm 1\label{Dirac}
\feq
holds, we can use the mass formula for so-called asymptotically magnetic AdS (mAdS) spacetimes,
obtained in \cite{Hristov:2011ye} for minimal gauged supergravity and generalized to include
matter-coupling in \cite{Hristov:2011qr}. Adapted to our conventions, it is given
by \cite{Hristov:2011qr}\footnote{In \cite{Hristov:2011qr}, the lower parts $F_I$ of the symplectic section are
(asymptotically) imaginary, whereas here they are real. These two choices are related by a global
$\text{U}(1)$ phase rotation (which is a symmetry of the theory). Notice also that the gauge coupling
constant $g$ of \cite{Hristov:2011qr} is two times our $g$, due to an additional factor of 4 in our scalar
potential w.r.t.~the one in \cite{Hristov:2011qr}. (Actually, $g$ does never appear explicitly here, since
it has been absorbed into the Fayet-Iliopoulos parameters $\xi_I$ by $g_I=g\xi_I$, cf.~\cite{Cacciatori:2009iz}).}
\eqn
M &=& \frac1{8\pi}\lim_{r\to\infty}\oint d\Sigma_{tr}\left(g'r + \frac{g'}{8g^2r}\right)\left(\pm2\text{Re}\,(Q_IX^I
            - P^IF_I)\sin\vartheta e^t_0 e^r_1 e^{\vartheta}_2 e^{\varphi}_3\right. \nonumber \\
    &+& \left. 4|g_IX^I| e^t_0 e^r_1 + (\omega^{12}_{\vartheta}e^t_0 e^r_1 e^{\vartheta}_2 +
            \omega^{13}_{\varphi}e^t_0 e^r_1 e^{\varphi}_3)\right)\,. \label{M-mAdS}
\feqn
Here we used a new radial coordinate $r=z/g'$, with $g'$ related to the asymptotic value of the cosmological
constant by $\Lambda=-3g'^2=-12\lim_{r\to\infty}g_Ig_J\bar X^IX^J$, which yields
\[
g' = 2\sqrt2 (g_0g_1g_2g_3)^{1/4}\,.
\]
Moreover, $t$ was rescaled, $t\to t/g'$ in order for the metric to be asymptotically mAdS in spherical
coordinates, cf.~\cite{Hristov:2011qr} for details. Note that the sign in the first line of \eqref{M-mAdS} is the
same as the one appearing in \eqref{Dirac}. The magnetic charges $P^I$ are to be identified with the
constants $C_{I+4}$, whereas the electric charges $Q_I$ vanish in our case. Evaluating \eqref{M-mAdS}
for the solution \eqref{metric-stu}-\eqref{F-stu}, one gets
\eqn
M &=& -\frac1{4\sqrt2}(g_0g_1g_2g_3)^{-1/4}[a_1 \pm 8\sum_I g_I^2\beta_{I+4}C_{I+4} \nonumber \\
       && \qquad + 96(g_1\beta_5 + g_2\beta_6)(g_0\beta_4 + g_1\beta_5)(g_0\beta_4 + g_2\beta_6)]\,,
             \label{M-mAdS-eval}
\feqn
where the sign in the first line corresponds again to the one in \eqref{Dirac}. The mass \eqref{M-mAdS}
appears on the rhs of the anticommutator of two supercharges \cite{Hristov:2011qr}, which implies
that the BPS bound for asymptotically mAdS spacetimes is given by $M=0$. As a consistency check,
let us verify that \eqref{M-mAdS-eval} vanishes in the supersymmetric case $a_1=0$,
$a_0=a_2^2/4$. \eqref{C_I+4}, together with \eqref{Dirac}, leads then to
\[
C_{I+4} = \pm\left(4g_I\beta_{I+4}^2 + \frac{a_2}{8g_I}\right)\,.
\]
Using this in \eqref{M-mAdS-eval}, one obtains indeed $M=0$. However, this is not the only zero
of $M$, and one might imagine for instance solutions with vanishing mass, but with $a_1\neq 0$,
which do not correspond to any known BPS black holes in matter-coupled gauged supergravity.
A similar behaviour was encountered in \cite{Toldo:2012ec} for the $t^3$ model (which is included in
our case), where it was argued that these solutions might be supersymmetric in extended supergravity
models \cite{Khuri:1995xk,Bena:2011pi}. It would be interesting to see whether this is actually the case.

In \cite{Hristov:2011qr}, an alternative notion of mass was proposed, namely
\eqn
M_{\text{hol}} &=& \frac1{8\pi}\lim_{r\to\infty}\oint d\Sigma_{tr}\left(e^t_{[0}e^r_1e^{\vartheta}_{2]}
                                  + \sin\vartheta e^t_{[0} e^r_1 e^{\varphi}_{3]}\right. \nonumber \\
    &+& \left. 4g'r|g_IX^I| e^t_{[0} e^r_{1]} + \sqrt{g'^2r^2+1}(\omega^{ab}_{\vartheta}e^t_{[0} e^r_a
            e^{\vartheta}_{b]} + \omega^{ab}_{\varphi}e^t_{[0} e^r_a e^{\varphi}_{b]})\right)\,. \label{M-hol}
\feqn
This expression, which does not require the charge quantization condition \eqref{Dirac} in order to
converge, is identical to the one coming from holographic renormalization. For the black hole
\eqref{metric-stu}-\eqref{F-stu}, one obtains
\eq
M_{\text{hol}} = -\frac1{4\sqrt2}(g_0g_1g_2g_3)^{-1/4}[a_1 + 96(g_1\beta_5 + g_2\beta_6)
                            (g_0\beta_4 + g_1\beta_5)(g_0\beta_4 + g_2\beta_6)]\,, \label{M-hol-eval}
\feq
which differs from \eqref{M-mAdS-eval} by the term linear in the charges. The holographic
renormalization procedure \cite{Balasubramanian:1999re,Bianchi:2001kw} consists in adding boundary counterterms to the supergravity action, that lead to finite conserved charges. It would be interesting
to see which finite counterterms have to be added in order to obtain the mass \eqref{M-mAdS} instead
of \eqref{M-hol}.

A final point we want to address in this subsection is the so-called area product formula.
It was found in \cite{Cvetic:2010mn} that for a large class of rotating multi-charge black holes in
asymptotically anti-de~Sitter spacetimes, the product of all horizon areas (including thus also inner
horizons) depends only on the charges, angular momenta and the cosmological constant.
In what follows, we will show that such universal results, which may provide a `looking glass' for
probing the microscopics of general black holes \cite{Cvetic:2010mn}, hold also for the solutions
constructed here\footnote{In the case of the $t^3$ model, the validity of an area product formula
was shown in \cite{Toldo:2012ec}.}. To this end, decompose $\exp(2\psi)=\prod_{i=1}^4(z-z_i)$.
Comparison with \eqref{quartic-pol} yields for the horizon locations $z_i$
\begin{eqnarray*}
&& a_0 = z_1z_2z_3z_4\,, \qquad a_1 = -(z_1 + z_2)z_3z_4 - (z_3 + z_4)z_1z_2\,, \\
&& a_2 = z_1z_2 + (z_1 + z_2)(z_3 + z_4) + z_3z_4\,, \qquad z_1 + z_2 + z_3 + z_4 = 0\,.
\end{eqnarray*}
Using this, it is easy to see that
\[
\prod_{i=1}^4 f_{I+4}(z_i) = \frac1{16g_I^2}\left[\frac{a_0}{16g_I^2} - \frac{a_1\beta_{I+4}}{4g_I} +
a_2\beta_{I+4}^2 + 16g_I^2\beta_{I+4}^4\right] = \frac{C^2_{I+4}}{16g_I^2}\,,
\]
where we used \eqref{C_I+4} in the last step. This gives for the product of the horizon areas
\eq
\prod_{i=1}^4 A(z_i) = \prod_i 2V(f_4f_5f_6f_7)^{1/2}|_{z_i} = \frac{V^4}{16g_0g_1g_2g_3}C_4C_5C_6C_7\,.
\feq
Taking into account that the asymptotic value of the cosmological constant is given by
$\Lambda=-3g'^2=-24(g_0g_1g_2g_4)^{1/2}$, this can be rewritten as
\eq
\prod_{i=1}^4 A(z_i) = \frac{36V^4}{\Lambda^2}C_4C_5C_6C_7\,, \label{area-prod}
\feq
which depends only on the charges and the asymptotic cosmological constant. Moreover, as
already noticed in \cite{Toldo:2012ec} for the $t^3$ model, the dependence on the charges
seems to be related to the square of the prepotential. Interestingly enough, the expression
\eqref{area-prod} coincides exactly with the one found in \cite{Cvetic:2010mn} for the
rotating black holes in ${\cal N}=4$ $\text{SO}(4)$ gauged supergravity with two pairwise equal
electric charges \cite{Chong:2004na}, if we set the angular momentum $J$ equal to zero, replace
$Q_1^2Q_2^2$ in \cite{Cvetic:2010mn} by $C_4C_5C_6C_7$ and take $V=4\pi$ in \eqref{area-prod}
(spherical case)\footnote{$\Lambda$ is related to the gauge coupling constant $g$
of \cite{Cvetic:2010mn} by $\Lambda=-3g^2$.}. This confirms the universality of the area product
formula.

\subsection{First-order equations and fake superpotential}

Note that the equations of motion \eqref{eq:eom1_q4}-\eqref{eq:eom1_q7}, together with \eqref{eq:eom1_E2},
follow from an action principle with Lagrangian
\eq
L = \sum_{I=0}^3\left[\frac14\dot\chi_I^2 + \frac14 C_{I+4}^2 e^{-2\chi_I} + 4g_I^2 e^{2\chi_I+4\psi}\right]
- \dot\psi^2 -4e^{4\psi}\left(\sum_{I=0}^3g_I e^{\chi_I}\right)^2 - \kappa e^{2\psi}\,. \label{1d-Lagrangian}
\feq
Here, a dot denotes a derivative w.r.t.~the coordinate $\zeta$ defined by $d\zeta=e^{-2\psi}dz$,
and we introduced $\chi_I=\ln q_{I+4}$. Eqn.~\eqref{eq:eom1_E1} does not follow from
\eqref{1d-Lagrangian}, but the linear combination \eqref{eq:eom1_E1}+\eqref{eq:eom1_E2} is nothing
else than the zero energy condition $H=0$, where $H$ is the Legendre transform of
\eqref{1d-Lagrangian}. If the potential $U$ appearing in \eqref{1d-Lagrangian} is given in terms
of a (fake) superpotential $W$,
\eq
-U = {\cal G}^{\alpha\beta}\frac{\partial W}{\partial\varphi^{\alpha}}\frac{\partial W}{\partial\varphi^{\beta}}\,,
\label{HamJac}
\feq
where $\alpha=(I,4)$, $\varphi^I=\chi_I$, $\varphi^4=\psi$ and
$({\cal G}^{\alpha\beta})=\text{diag}(4,4,4,4,-1)$, one can rewrite the Lagrangian (up to total derivatives)
in the form
\eq
L = {\cal G}_{\alpha\beta}\left(\dot\varphi^{\alpha} - {\cal G}^{\alpha\gamma}\frac{\partial W}
{\partial\varphi^{\gamma}}\right)\left(\dot\varphi^{\beta} - {\cal G}^{\beta\delta}\frac{\partial W}
{\partial\varphi^{\delta}}\right)\,,
\feq
and thus the action is stationary if the first-order equations
\eq
\dot\varphi^{\alpha} = {\cal G}^{\alpha\beta}\frac{\partial W}{\partial\varphi^{\beta}} \label{first-order}
\feq
hold. Notice that \eqref{HamJac} is just the reduced Hamilton-Jacobi equation in the case of zero
`energy', with Hamilton's characteristic function $W$\footnote{For further discussions of the
relationship between the Hamilton-Jacobi formalism and the first-order equations derived from a fake
superpotential cf.~\cite{Andrianopoli:2009je,Trigiante:2012eb}.}.

To solve \eqref{HamJac}, we make the ansatz
\eq
W = \sum_{I=0}^3(a_I e^{-\chi_I} + b_I e^{\chi_I+2\psi})\,. \label{ansatz-W}
\feq
Then, \eqref{HamJac} is satisfied if the following conditions hold:
\eq
a_I = \frac14C_{I+4}\epsilon_I\,, \qquad b_I = g_I\,, \qquad \kappa = 2\sum_I g_IC_{I+4}\epsilon_I\,,
\label{abkappa}
\feq
where $\epsilon_I=\pm 1$ are arbitrary signs. The last eqn.~of \eqref{abkappa} represents a constraint
on the magnetic charges, and thus \eqref{ansatz-W} does not solve \eqref{HamJac} for arbitrary values
of $C_{I+4}$. Using \eqref{ansatz-W} and \eqref{abkappa} in \eqref{first-order} gives the first-order
equations
\eqn
\partial_z q_{I+4} &=& 4g_Iq_{I+4}^2 - \epsilon_IC_{I+4}e^{-2\psi}\,, \quad \text{(no sum over $I$)} \nonumber \\
\partial_z\psi &=& -2\sum_I g_I q_{I+4}\,,
\feqn
which are solved by
\[
q_{I+4} = \frac{\alpha_Iz+\beta_I}{e^{\psi}}\,, \qquad e^{\psi} = z^2 + c\,,
\]
with the constants $\alpha_I,\beta_I,c$ satisfying\footnote{Here, $\alpha_I$ is negative while
$\alpha_{I+4}$ in \eqref{ansatz-q} is positive. $\alpha_I>0$ can easily be achieved by taking $z\to-z$.}
\eq
\alpha_I = -\frac1{4g_I}\,, \qquad \sum_I g_I\beta_I = 0\,, \qquad c + 16g_I^2\beta_I^2 = 4g_I\epsilon_I
C_{I+4}\,. \label{alpha-beta-c}
\feq
Summing over $I$ in the last eqn.~and using \eqref{abkappa} yields
\eq
c = \frac{\kappa}2 - 4\sum_Ig_I^2\beta_I^2\,. \label{c}
\feq
We have thus a three-parameter solution (four $\beta_I$'s minus the constraint $\sum_Ig_I\beta_I=0$;
\eqref{c} determines then $c$ and \eqref{alpha-beta-c} gives the magnetic charges). Apart from these
three continuous parameters, the solution is labeled also by the five discrete constants $\kappa$ and
$\epsilon_I$. In order to have a horizon we need $c<0$ and thus
\eq
8\sum_Ig_I^2\beta_I^2 > \kappa\,. \label{outside-ball}
\feq
This does always hold for $\kappa=0,-1$ (flat or hyperbolic horizon). For $\kappa=1$ the allowed values
of the $\beta_I$ lie on the intersection of a plane through the origin (second eqn.~of \eqref{alpha-beta-c})
with the exterior of a ball in $\mathbb{R}^4$, eqn.~\eqref{outside-ball}. Since $e^{2\psi}$ has a double root,
the resulting black hole is extremal. If all signs $\epsilon_I$ are equal, it reduces to the supersymmetric
solution found in \cite{Cacciatori:2009iz}.

\subsection{Recovering the four-charge solution of ${\cal N}=8$ gauged supergravity}

Let us now show how the nonextremal black holes of section \ref{nonextr-bh} contain the magnetic
four-charge solution to ${\cal N}=8$ gauged supergravity presented in \cite{Duff:1999gh}.
After a $\text{U}(1)^4$ truncation, this theory boils down to the ${\cal N}=2$ model considered here,
with prepotential $F=-2i(X^0X^1X^2X^3)^{1/2}$, and all $g_I$ equal. The metric, moduli and gauge
fields read (cf.~eqn.~(6.2) of \cite{Duff:1999gh})
\begin{eqnarray}
&& ds^2 = -(H_0H_1H_2H_3)^{-1/2}f dt^2 + (H_0H_1H_2H_3)^{1/2}\left(\frac{dr^2}f + r^2d\Omega^2
                    \right)\,, \nonumber \\
&& e^{2\phi^{(12)}} = \frac{H_2H_3}{H_0H_1}\,, \qquad e^{2\phi^{(13)}} = \frac{H_1H_3}{H_0H_2}\,,
\qquad e^{2\phi^{(14)}} = \frac{H_1H_2}{H_0H_3}\,, \nonumber \\
&& H_I = 1 + \frac{k\sinh^2\!\mu_I}r\,, \qquad f = 1 - \frac kr + 2g^2r^2 H_0H_1H_2H_3\,, \nonumber \\
&& F^I_{\vartheta\varphi} = \frac{\eta_I}{2\sqrt2}k\cosh\mu_I\sinh\mu_I\sin\vartheta\,. \label{Duff-sol}
\end{eqnarray}
Here, $\eta_I=\pm 1$ are arbitrary signs, the $\mu_I$ determine the magnetic charges, and $k$ is a kind of nonextremality parameter. After the identification $g=2g_I$ and the coordinate transformation
\[
r = \frac z{\sqrt2 g} - \frac k4\sum_{I=0}^3\sinh^2\!\mu_I\,,
\]
the solution \eqref{Duff-sol} takes the form of the one in section \ref{nonextr-bh}, with
\eqn
&& \beta_{I+4} = \frac k{\sqrt2}\left(\sinh^2\!\mu_I - \frac14\sum_J\sinh^2\!\mu_J\right)\,, \nonumber \\
&& a_0 = \frac{g^2k^2}2\left[\left(\frac12\sum_I\sinh^2\!\mu_I\right)^2 + \sum_I\sinh^2\!\mu_I\right]
                 + 16g^4\beta_4\beta_5\beta_6\beta_7\,, \nonumber \\
&& a_1 = -\sqrt2 gk\left(1 + \frac12\sum_I\sinh^2\!\mu_I\right) - \sqrt2\frac{g^3k^3}4\left(\sinh^2\!\mu_0 +
                  \sinh^2\!\mu_1 - \sinh^2\!\mu_2\right. \nonumber \\
&&             \qquad\left.- \sinh^2\!\mu_3\right)\left[(\sinh^2\!\mu_2 - \sinh^2\!\mu_3)^2 - (\sinh^2\!\mu_0 -
                  \sinh^2\!\mu_1)^2\right]\,, \nonumber \\
&& a_2 = 1 - 2g^2\sum_I\beta_{I+4}^2\,. \label{parametr-Duff}
\feqn
Since both \eqref{Duff-sol} and the black holes obtained in \ref{nonextr-bh} are labeled by five
continuous parameters ($k,\mu_I$ for \eqref{Duff-sol}), one might wonder if they are
not equivalent (if all $g_I$ are equal; for generic coupling constants the solution
\eqref{metric-stu}-\eqref{F-stu} is clearly more general). This is however not the case: Suppose for
instance that all charges are equal in \eqref{Duff-sol}. Then, the `harmonic' functions $H_I$ coincide
as well, and thus the scalar fields are constant. In the solution of section \ref{nonextr-bh} instead,
one can have equal charges and yet nontrivial profiles for the moduli (take e.g.~$a_1=0$,
$\beta_4=\beta_5=-\beta_6=-\beta_7$). Moreover, \eqref{metric-stu}-\eqref{F-stu} contains a subclass
of black holes that are BPS, while it was shown in \cite{Duff:1999gh} that \eqref{Duff-sol} can never
be supersymmetric. To understand better what happens, let us consider the subcase $\beta_6=\beta_4$,
$\beta_7=\beta_5$, $\mu_2=\mu_0$, $\mu_3=\mu_1$, that implies $X^2=X^0$, $X^3=X^1$, $C_6=C_4$, $C_7=C_5$, leading to the $F=-iX^0X^1$ model considered in \cite{Klemm:2012yg}.
Since all $g_I$ are equal ($2g_I=g$), the second equ.~of \eqref{alpha-beta-a2} boils down to
$\sum_I\beta_{I+4}=0$ and thus $\beta_5=-\beta_4$. From \eqref{C_I+4} one obtains then
\[
C_5^2 - C_4^2 = \frac{a_1}g\beta_4\,.
\]
If the charges are equal (up to a sign), $C_5^2=C_4^2$, and we have therefore $a_1=0$ or $\beta_4=0$.
The former case is (for $C_5=C_4$) the supersymmetric black hole found in
\cite{Cacciatori:2009iz} (with running scalar), whereas the latter corresponds to the Duff-Liu solution
\eqref{Duff-sol}, with constant scalar profiles. In this context, notice also that in the parametrization
\eqref{parametr-Duff}, for $\mu_2=\mu_0$, $\mu_3=\mu_1$, we get
\[
a_1 = -\sqrt2 gk(1 + \sinh^2\!\mu_0 + \sinh^2\!\mu_1)\,,
\]
which is always nonvanishing (if $k\neq 0$), and thus the supersymmetric case cannot appear.

In conclusion, the solution \eqref{metric-stu}-\eqref{F-stu} includes both \eqref{Duff-sol} and
the BPS black holes constructed in \cite{Cacciatori:2009iz}.

\section{The $F(X) = -X^1X^2X^3/X^0$ model}
\label{-X1X2X3/X0-model}

In \cite{Mohaupt:2011aa} the Hesse potential for the $F(X) = -X^1X^2X^3/X^0$ model was calculated to be
	\begin{align*}
			H &= - 2 \Big( -(y_0x^0 + y_1x^1 + y_2 x^2 + y_3 x^3)^2 + 4y_1 x^1 y_2 x^2 + 4y_1 x^1 y_3 x^3 \\
				&\hspace{12em} + 4 y_2 x^2 y_3 x^3	+ 4 x^0 y_1 y_2 y_3 - 4 y_0 x^1 x^2 x^3 \Big)^{1/2} \;. 
	\end{align*}

\subsection{Equations of motion}

	We will consider solutions which take the form
	\begin{equation}
		y_0 = x^1 = x^2 = x^3 = 0 \qquad \Rightarrow \qquad u^0 = v_1 = v_2 = v_3 = 0 \;. \label{eq:restriction2}
	\end{equation}
	This ensures that the kinetic term in the Lagranian does not transform under any axion-like shift symetries of the physical scalar fields.
	The matrix $\tilde{H}^{ab}$ is given by
		\begin{equation}
		\tilde{H}^{ab} = \left( \begin{array}{cccccccc} 
		4{x^0}^2	& 0 & 0 & 0 & 0 & 0 & 0 & 0 \\
		0	& * & * & * & * & 0 & 0 & 0 \\
		0	& * & * & * & * & 0 & 0 & 0 \\
		0	& * & * & * & * & 0 & 0 & 0 \\
		0 & * & * & * & * & 0 & 0 & 0 \\
		0 & 0 & 0 & 0 & 0 & 4{y_1}^2 & 0 & 0 \\		
		0 & 0 & 0 & 0 & 0 & 0 & 4{y_2}^2 & 0 \\		
		0 & 0 & 0 & 0 & 0 & 0 & 0 & 4{y_3}^2		
		\end{array} \right) \;. \label{eq:Htilde2}
	\end{equation}
	The equations of motion greatly simplify if we impose that in addition half of the harmonic functions are constant,
	\[
		\partial_\mu {\cal H}_1 = \partial_\mu {\cal H}_2 = \partial_\mu {\cal H}_3 = \partial_\mu {\cal H}_4 = 0 \;. 
	\]
	Using the decomposition of $\tilde{H}^{ab}$ and (\ref{eg:qhat}) one can see that this condition is equivalent to switching off the electric charges $Q_1,Q_2,Q_3$ and the magentic charge $P^1$. Moreover, it implies that block in $\tilde{H}^{ab}$ consisting of entries denoted by $*$ completely decouples from the equations of motion.
	The non-vanishing dual coordinates read
	\begin{align}
		q_0 &= -\frac{1}{4x^0} \;, & q_6 &= -\frac{1}{4y_2} \;, \notag \\
		q_5 &= -\frac{1}{4y_1} \;, & q_7 &= -\frac{1}{4y_3} \;. \label{eq:coords_and_duals2}
	\end{align}
	The second order equations of motion \eqref{eq:eom_dual1} for the $q_a$ are given by
	\begin{align}
		&\Delta {q}_0 - \frac{\left[ (\partial_z q_0)^2 - (\partial_z {\cal H}_0)^2 \right]}{q_0}  = 0 \;, \label{eq:eom2_q0} \\
		&\Delta {q}_5 - \frac{\left[ (\partial_z q_5)^2 - (\partial_z {\cal H}_5)^2 \right]}{q_5} \notag \\
		&\hspace{6em} + 16g_1 q_5^2 \left(g_2 q_6 + g_3 q_7 \right) = 0 \;, \label{eq:eom2_q5} \\ 
		&\Delta {q}_6 - \frac{\left[ (\partial_z q_6)^2 - (\partial_z {\cal H}_6)^2 \right]}{q_6} \notag \\
		&\hspace{6em} + 16g_2 q_6^2 \left(g_1 q_5 + g_3 q_7 \right) = 0 \;, \label{eq:eom2_q6} \\
		&\Delta {q}_7 - \frac{\left[ (\partial_z q_7)^2 - (\partial_z {\cal H}_7)^2 \right]}{q_7} \notag \\
		&\hspace{6em} + 16g_3 q_7^2 \left(g_1 q_5 + g_2 q_6\right) = 0 \;, \label{eq:eom2_q7}
	\end{align}
	and the Einstein equations \eqref{eq:eom_dual2} become
	\begin{align}
		& \frac{\left[ (\partial_z q_0)^2 - (\partial_z {\cal H}_0)^2 \right]}{4q_0^2} + 
		\frac{\left[(\partial_z q_5)^2 - (\partial_z {\cal H}_5)^2 \right]}{4 q_5^2}  \notag \\
		&+ \frac{\left[(\partial_z q_6)^2 - (\partial_z {\cal H}_6)^2 \right]}{4 q_6^2}  +
		\frac{\left[(\partial_z q_7)^2 - (\partial_z {\cal H}_7)^2 \right]}{4 q_7^2} \notag \\
		&\hspace{6em} - 8g_1 g_2 q_5 q_6 - 8 g_1 g_3 q_5 q_7 - 8g_2g_3 q_6 q_7 = -\partial_z^2 \psi - (\partial_z \psi)^2 \;, \label{eq:eom2_E1} \\
		\notag \\
		& 16g_1 g_2 q_5 q_6 + 16 g_1 g_3 q_5 q_7 + 16g_2g_3 q_6 q_7  = \partial^2_z \psi +2 (\partial_z \psi)^2 - {\kappa} e^{-2\psi} \;. \label{eq:eom2_E2}
	\end{align}	
	One can check that upon making the truncation
	\begin{align*}
		&q_0 \to q_0 \;, & &g_0 \to g_0 \;, \\
		&q_5 \to \tfrac13 q_3 \;, & &g_1 \to g_1 \;, \\
		&q_6 \to \tfrac13 q_3 \;, & &g_2 \to g_1 \;, \\
		&q_7 \to \tfrac13 q_3 \;, & &g_3 \to g_1 \;,
	\end{align*}
	one obtains the equations of motion for the $F = - {X^1}^3 / X^0$ model found in \cite{Klemm:2012yg}.

Note that the fields ${\cal H}_0, {\cal H}_5, {\cal H}_6, {\cal H}_7$ are harmonic functions, i.e.
\[
\partial_z {\cal H}_{0} = -D e^{-2\psi} \;, \;\;\;\; 
\partial_z {\cal H}_{5} = C_5 e^{-2\psi} \;, \;\;\;\;
\partial_z {\cal H}_{6} = C_6 e^{-2\psi} \;, \;\;\;\;
\partial_z {\cal H}_{7} = C_7 e^{-2\psi} \;.
\]
where the $C_{5,6,7}$ are constants proportional to the magnetic charges and $D$ is a constant proportional to the electric charge.

\subsection{First-order equations and fake superpotential}

As was explained in \cite{Klemm:2012yg}, \eqref{eq:eom2_q0} is nothing else than the Liouville
equation, with the solution
\eq
q_0 = D\frac{\sin p\zeta}p\,, \label{q_0}
\feq
where again $d\zeta=e^{-2\psi}dz$, and $p$ (with $p^2$ real) is the Liouville momentum. Using this,
the Einstein equation \eqref{eq:eom2_E1} boils down to
\begin{eqnarray}
&&\frac{p^2}4e^{-4\psi} + \sum_{\alpha=5}^7\frac{(\partial_z q_{\alpha})^2 - C_{\alpha}^2 e^{-4\psi}}
{4q_{\alpha}^2} - 8g_1g_2q_5q_6 - 8g_1g_3q_5q_7 - 8g_2g_3q_6q_7 \nonumber \\
&& \qquad = -\partial^2_z\psi
- (\partial_z\psi)^2\,, \label{eq:eom2_E1prime}
\end{eqnarray}
and thus $q_0$ decouples completely from the other fields.

The equations of motion \eqref{eq:eom2_q5}-\eqref{eq:eom2_q7}, together with \eqref{eq:eom2_E2},
follow from an action principle with the Toda-like Lagrangian
\begin{eqnarray}
\lefteqn{L = -\dot\psi^2 + \frac14\sum_{\alpha=5}^7((\dot\chi^{\alpha})^2 + C_{\alpha}^2e^{-2\chi^{\alpha}})}
\nonumber \\
&& -8 e^{4\psi}(g_1g_2 e^{\chi^5 + \chi^6} + g_1g_3 e^{\chi^5 + \chi^7} + g_2g_3 e^{\chi^6 + \chi^7})
-\kappa e^{2\psi}\,, \label{L-other-stu}
\end{eqnarray}
where a dot denotes a derivative w.r.t.~$\zeta$, and we defined $\chi^{\alpha}\equiv\ln q_{\alpha}$.
Notice that \eqref{eq:eom2_E1prime} does not follow from \eqref{L-other-stu}, but the linear combination
\eqref{eq:eom2_E1prime}+\eqref{eq:eom2_E2} is exactly the `energy conservation' $H=-p^2/4\equiv E$,
with $H$ the Legendre transform of $L$. It is amusing to note that the `energy' is essentially the square
of the Liouville momentum.

Defining the metric $\cal G$ by $({\cal G}^{\alpha\beta})=\text{diag}(4,4,4)$, ${\cal G}^{\psi\psi}=-1$,
${\cal G}^{\alpha\psi}=0$, one can write
\eq
L = {\cal G}_{\alpha\beta}\dot\chi^{\alpha}\dot\chi^{\beta} + {\cal G}_{\psi\psi}\dot\psi^2 - U\,,
\feq
with $U$ the potential term appearing in \eqref{L-other-stu}. Then, the reduced Hamilton-Jacobi equation
reads
\eq
{\cal G}^{\alpha\beta}\frac{\partial W}{\partial\chi^{\alpha}}\frac{\partial W}{\partial\chi^{\beta}}
+ {\cal G}^{\psi\psi}\left(\frac{\partial W}{\partial\psi}\right)^2 + U = E\,, \label{HJ-other-stu}
\feq
and we have the first-order equations
\eq
\dot\chi^{\alpha} = {\cal G}^{\alpha\beta}\frac{\partial W}{\partial\chi^{\beta}}\,, \qquad \dot\psi =
{\cal G}^{\psi\psi}\frac{\partial W}{\partial\psi}\,. \label{1st-order-other-stu}
\feq
If $E=0$ (i.e., $p=0$) and the generalized `charge quantization condition'
\eq
\sum_{\alpha=5}^7 g_{\alpha-4}\epsilon_{\alpha}C_{\alpha} = \frac{\kappa}2 
\feq
holds, where $\epsilon_{\alpha}=\pm 1$ are arbitrary signs, a particular solution to \eqref{HJ-other-stu}
is given by
\eq
W = \sum_{\alpha=5}^7\left(\frac{\epsilon_{\alpha}C_{\alpha}}4 e^{-\chi^{\alpha}} + g_{\alpha-4}
e^{\chi^{\alpha} + 2\psi}\right)\,, \label{W-other-stu}
\feq
which has exactly the same form as the fake superpotential \eqref{ansatz-W} for the model considered
in section \ref{sqrtX0X1X2X3-model}, the only (but crucial) difference being that the sum in
\eqref{ansatz-W} involves four terms, whereas in \eqref{W-other-stu} one has only three\footnote{We shall
see below that this affects the asymptotics of the resulting black hole.}.
There seems thus to be a universal structure behind the form of the fake superpotential. We shall come
back to this point later.

With \eqref{W-other-stu}, the first-order equations \eqref{1st-order-other-stu} become
\eq
\partial_z\psi = -2\sum_{\alpha}g_{\alpha-4}q_{\alpha}\,, \qquad \partial_z q_{\alpha} = 4g_{\alpha-4}
q_{\alpha}^2 - \epsilon_{\alpha}C_{\alpha}e^{-2\psi}\,, \label{1st-order-other-stu-prime}
\feq
which differ from the BPS equations (3.69) and (3.70) of \cite{Cacciatori:2009iz} only by the
signs $\epsilon_{\alpha}$\footnote{The dictionary to get from \eqref{1st-order-other-stu-prime} (with
$\epsilon_{\alpha}=-1$) to (3.69) and (3.70) of \cite{Cacciatori:2009iz} is $q_{\alpha}=2iH^{\alpha-4}$,
$C_{\alpha}=4\pi p^{\alpha-4}$.}. For $\epsilon_{\alpha}$ all equal one recovers the BPS case.

A special solution of \eqref{1st-order-other-stu-prime} is
\eq
e^{\psi} = z\,, \qquad q_{\alpha} = \frac{\pi_{\alpha}}z\,,
\feq
where the constants $\pi_{\alpha}$ satisfy
\eq
4g_{\alpha-4}\pi_{\alpha}^2 + \pi_{\alpha} - \epsilon_{\alpha}C_{\alpha} = 0\,.
\feq
Note that the equations
\eqref{additional-ansatz}, \eqref{static} (with $\tilde{\phi}=0$) are trivially satisfied in this case.
The dilaton $\phi$ is computed from the Hesse potential,
\[
e^{\phi} = -2H(x,y) = \frac12 (q_0q_5q_6q_7)^{-1/2}\,.
\]
This yields for the four-dimensional metric
\begin{equation}
ds_4^2 = -\frac{z^{3/2} dt^2}{2(q_0\pi_5\pi_6\pi_7)^{1/2}} + \frac{2(q_0\pi_5\pi_6
\pi_7)^{1/2}}{z^{3/2}}\left(dz^2 + z^2(d\vartheta^2 + S_{\kappa}^2(\vartheta)d\varphi^2)\right)\,.
\label{metr-other-stu}
\end{equation}
Here, $q_0$ is determined by \eqref{q_0} (in the limit $p\to 0$), which gives
\[
q_0 = h_0 - \frac Dz\,,
\]
where $h_0$ is an integration constant, and the coordinates $\vartheta,\varphi$ as well as the
function $S_{\kappa}(\vartheta)$ were defined in section \ref{nonextr-bh}.

The solution \eqref{metr-other-stu} has an event horizon at $z=0$. The near-horizon geometry is
$\text{AdS}_2\times\Sigma$, where $\Sigma$ is a two-space of constant curvature. Note that,
for $z\to\infty$, the spacetime does not approach AdS. In the special case $\epsilon_{\alpha}=-1$,
$g_1C_5=g_2C_6=g_3C_7$, $g_1q_5=g_2q_6=g_3q_7$ one recovers the BPS black hole
(3.81) of \cite{Cacciatori:2009iz} with $C=0$.

The scalar fields follow from \eqref{xy}, with the result
\eq
X^0 = -\frac{(q_0q_5q_6q_7)^{1/4}}{2\sqrt2 q_0}\,, \qquad
X^{\alpha-4} = -\frac{iq_{\alpha}}{2\sqrt2(q_0q_5q_6q_7)^{1/4}}\,.
\feq
Finally, from (\ref{eq:field_strengths}) the nonvanishing components of the gauge field strengths read
\[
F^0_{tz} = -\frac12\partial_z(q_0^{-1})\,, \qquad G_{\alpha-4|tz} = -\frac{C_{\alpha}}{2\pi_{\alpha}^2}\,,
\]
and using the fact that
\[
{\cal N}^{-1} = i(q_0q_5q_6q_7)^{-1/2}\text{diag}(q_0^{-1}q_5q_6q_7,q_5^2,q_6^2,q_7^2)\,,
\]
we can write this as
\eq
F^0 = -\frac12 dt\wedge d(q_0^{-1})\,, \qquad F^{\alpha-4} = \frac i2 C_{\alpha}e^{2\gamma}
dw\wedge d\bar w\,. \label{F-other-stu}
\feq
Observe that the expressions for the gauge field strengths are precisely the same as for the BPS
case \cite{Cacciatori:2009iz}. Since the black hole \eqref{metr-other-stu} is extremal, its
temperature is zero. For the Bekenstein-Hawking entropy one obtains
\begin{equation}
S = \frac{A_{\text h}}{4G} = \frac{(-D\pi_5\pi_6\pi_7)^{1/2}V}{2G}\,,
\end{equation}
where $V$ was defined in \eqref{def-vol}.

\section{Fake superpotential for arbitrary prepotential}
\label{W-arbitr-prepot}

For the three-dimensional base space metric given in section \ref{sec-metr-ans}, the equations of
motion \eqref{eq:eom_dual1}, \eqref{eq:eom_dual2} reduce to
\begin{eqnarray}
&&e^{-2\psi}\partial_z(e^{2\psi}\partial_z q_a) + \frac12\partial_a\tilde H^{bc}(\partial_z q_b\partial_z q_c
- C_a C_b e^{-4\psi}) + \partial_a\left(\frac{V(q)}{4H}\right) = 0\,, \label{eom_scal} \\
&& \tilde H^{ab}(\partial_z q_a\partial_z q_b - C_a C_b e^{-4\psi}) - \frac1{2H}V(q) = -\partial_z^2\psi
-(\partial_z\psi)^2\,, \label{eom_Einst1} \\
&&\frac1H V(q) = \partial_z^2\psi + 2(\partial_z\psi)^2 - \kappa e^{-2\psi}\,, \label{eom_Einst2}
\end{eqnarray}
where we took into account that the ${\cal H}_a$ are harmonic functions on the base, i.e.,
$\partial_z{\cal H}_a = C_a e^{-2\psi}$, with charges $C_a$. Eqns.~\eqref{eom_scal} and
\eqref{eom_Einst2} follow from an action principle with Lagrangian
\eq
L = \tilde H^{ab}\dot q_a\dot q_b - \dot\psi^2 + \tilde H^{ab}C_a C_b - \frac{V}{2H}e^{4\psi} - \kappa e^{2\psi}\,,
\label{general-Lagr}
\feq
where, as before, a dot denotes a derivative w.r.t.~the coordinate $\zeta$ defined by $d\zeta=e^{-2\psi}dz$.
Notice that \eqref{eom_Einst1} does not follow from \eqref{general-Lagr}, but the linear combination
\eqref{eom_Einst1}+\eqref{eom_Einst2} is precisely the `zero-energy condition' $H=0$, with $H$ the
Legendre transform of $L$. Let us define the index $\alpha=(a,2n+2)$ (i.e., $\alpha=0,\ldots,2n+2$),
the fields $\varphi_{\alpha}$ with $\varphi_a=q_a$, $\varphi_{2n+2}=\psi$, and the `metric'
\[
\left({\cal G}_{\alpha\beta}\right) = \left(\begin{array}{cc} \tilde H_{ab} & 0 \\ 0 & -1\end{array}\right)\,.
\]
If the potential $U$ appearing in \eqref{general-Lagr} is given in terms of
a fake superpotential $W$,
\eq
- U = {\cal G}_{\alpha\beta}\frac{\partial W}{\partial\varphi_{\alpha}}\frac{\partial W}{\partial\varphi_{\beta}}\,,
\label{Ham-Jac-general}
\feq
one can rewrite the Lagrangian $L$ (up to total derivatives) in the form
\eq
L = {\cal G}^{\alpha\beta}\left(\dot\varphi_{\alpha} - {\cal G}_{\alpha\gamma}\frac{\partial W}
{\partial\varphi_{\gamma}}\right)\left(\dot\varphi_{\beta} - {\cal G}_{\beta\delta}\frac{\partial W}
{\partial\varphi_{\delta}}\right)\,,
\feq
and thus the action is stationary if the first-order equations
\eq
\dot\varphi_{\alpha} = {\cal G}_{\alpha\gamma}\frac{\partial W}{\partial\varphi_{\gamma}}
\label{flow-eqns-gen}
\feq
hold. Again, \eqref{Ham-Jac-general} is just the reduced Hamilton-Jacobi equation in the case of zero
`energy', with Hamilton's characteristic function $W$. Using the expression \eqref{eq:potential_H}
for the Fayet-Iliopoulos potential in terms of the Hesse potential, and imposing the zero-axion
condition $x^I=0$, it is straightforward to show that a particular solution to \eqref{Ham-Jac-general}
is given by
\eq
W = q^a{S_a}^bC_b + k^aq_a e^{2\psi}\,, \label{W-general}
\feq
where $k^a=(0,\ldots,0,\pm g_I)^T$, and ${S_a}^b$ denotes a constant `field rotation matrix' that must
satisfy the compatibility condition
\eq
\tilde H^{cd} = \tilde H^{ab}{S_a}^c{S_b}^d \label{comp-cond}
\feq
with $\tilde H^{ab}$. Moreover, in order for \eqref{W-general} to satisfy \eqref{Ham-Jac-general},
the Dirac-type charge quantization condition
\eq
2k^a{S_a}^bC_b = -\kappa \label{Dirac-gen}
\feq
must hold. A nontrivial $S$ (different from $\pm\text{Id}$) allows to generate new solutions
from known ones by `rotating charges'. This technique was first introduced in
\cite{Ceresole:2007wx,LopesCardoso:2007ky}, and generalizes the sign-flipping procedure
of \cite{Ortin:1996bz}, which was used in the preceding two sections. Note that in general it is not
guaranteed that a nontrivial field rotation matrix satisfying \eqref{comp-cond} exists. Geometrically,
this is equivalent to the problem of identifying totally geodesic, totally isotropic submanifolds of the
scalar target space \cite{Mohaupt:2011aa}. For instance, it was established in \cite{Mohaupt:2011aa}
that a nontrivial solution to \eqref{comp-cond} always exists for non-axionic solutions of models with
a prepotential of the form $F=f(Y^1,\ldots,Y^n)/Y^0$, where $f$ is real when evaluated on real fields.
Below we will show that $S=\pm\text{Id}$ leads precisely to the BPS solutions of \cite{Cacciatori:2009iz}.

Before we proceed, a short comment on the ungauged limit is in order. In this case, $\kappa=k^a=0$
and thus \eqref{Dirac-gen} holds identically. The (fake) superpotential reduces to
$W=q^a{S_a}^bC_b$. If we restrict to BPS solutions, we can take without loss of generality
${S_a}^b={\delta_a}^b$ (the other sign corresponds just to a redefiniton $z\to-z$). Since $\psi=0$,
the Hamilton-Jacobi equations boil down to $\partial_z q_a = \tilde H_{ab}\partial_{q_b}W=C_a$.
On-shell one has therefore $W=q^a\partial_z q_a$. Taking into account that $q^aq_a=-1$ and
$q_a=\partial_a\tilde H$, we get thus $W=-\partial_z\tilde H$, so Hamilton's characteristic function
is just minus the derivative of the Hesse potential.

Coming back to the gauged case, the flow equations \eqref{flow-eqns-gen} read
\eq
\dot q_a = {S_a}^bC_b + \tilde H_{ab}k^b e^{2\psi}\,, \label{flow-q_a}
\feq
\eq
\dot\psi = -2k^a q_a e^{2\psi} = \mp 2g_I\frac{\partial\tilde H}{\partial y_I}e^{2\psi}\,. \label{flow-psi}
\feq
Of course, these have to be supplemented by the constraints \eqref{additional-ansatz} and the integrability
condition \eqref{static}. While the first equ.~of \eqref{additional-ansatz} holds automatically provided
$x^I=0$, the second equ.~of \eqref{additional-ansatz} and \eqref{static} become respectively
\eq
q^a\Omega_{ab}\tilde H^{bc}C_c = 0\,, \qquad \hat q^a\Omega_{ab}\tilde H^{bc}C_c = 0\,.
\label{additional-constr-gen}
\feq
Using $\psi$ in place of $z$ as a radial coordinate, we can finally write for the metric\footnote{Note that
$H<0$.}
\eq
ds_4^2 = 2Hdt^2 - \frac1{2H}\left[\left(2g_I\frac{\partial\tilde H}{\partial y_I}\right)^{\!\!-2}\!d\psi^2
                 + e^{2\psi}(d\vartheta^2 + S_{\kappa}^2(\vartheta)d\varphi^2)\right]\,, \label{metric-gen}
\feq
while the scalars satisfy the flow equation
\eq
\frac{dq_a}{d\psi} = -\frac1{2k^cq_c}\left({S_a}^bC_b e^{-2\psi} + \tilde H_{ab}k^b\right)\,.
\feq
In order to specify the full solution, one also has to give the expression for the fluxes, which we did
not write down here.

Let us now show that for $S=\pm\text{Id}$, one recovers the supersymmetric black holes constructed
in \cite{Cacciatori:2009iz}. First of all, it is easy to see that \eqref{flow-psi} corresponds exactly to
equ.~(2.22) of \cite{Cacciatori:2009iz} if we take the upper sign and identify the field $b$ used in
\cite{Cacciatori:2009iz} with $e^{\phi/2}$. Moreover, taking into account that
\[
\tilde H_{ab} = \frac1H\hat H_{ab} - \frac2{H^2}\Omega_{ac}q^c\Omega_{bd}q^d\,,
\]
as well as
\[
\hat H_{ab} = \left(\begin{array}{c@{\quad}c} \text{Im}\,{\cal N} + \text{Re}\,{\cal N}(\text{Im}\,{\cal N})^{-1}
                        \text{Re}\,{\cal N} & -\text{Re}\,{\cal N}(\text{Im}\,{\cal N})^{-1} \\
                        -(\text{Im}\,{\cal N})^{-1}\text{Re}\,{\cal N} & (\text{Im}\,{\cal N})^{-1} \end{array}\right)\,,
\]
together with $k^a\Omega_{ab}q^b=0$ for non-axionic solutions and $C_a\equiv 4\pi(q_I,-p^I)^T$,
one may check that \eqref{flow-q_a} is equivalent to eqns.~(2.20) and (2.21) of \cite{Cacciatori:2009iz}
for ${S_a}^b=-{\delta_a}^b$. Finally, the staticity condition (2.19) of \cite{Cacciatori:2009iz} can be
rewritten as $\dot q_a\Omega^{ab}q_b=0$, which is also satisfied by virtue of \eqref{flow-q_a},
$C_a\Omega^{ab}q_b=0$ and $k^a\tilde H_{ab}\Omega^{bc}q_c=0$. Note that the last two equations
can be shown by using $H_{ab}\Omega^{bc}H_{cd}=-4\Omega_{ad}$,
\[
\tilde H_{ab} = -\frac1{2H}H_{ab} + \frac1{2H^2}H_a H_b\,,
\]
as well as the first relation of \eqref{additional-constr-gen}.

We now wish to compare the fake superpotential \eqref{W-general} with results
that appeared previously in the literature. The authors of \cite{Dall'Agata:2010gj}\footnote{See also
\cite{Barisch:2011ui} for related work in the case of brane solutions ($\kappa=0$). In five-dimensional
gauged supergravity, the formalism of first-order flow equations was used in \cite{Cardoso:2008gm}
and more recently for extremal black branes in \cite{BarischDick:2012gj}.} rewrote the
supergravity action (reduced to one dimension) as a sum of squares of first-order differential conditions.
This was done for a generic prepotential, but was possible only provided a certain supersymmetry constraint
((2.27) of \cite{Dall'Agata:2010gj}, corresponds essentially to \eqref{Dirac-gen} with
$S=\pm\text{Id}$) holds. The resulting first-order equations involve the `superpotential'\footnote{The
function $U$ used in \cite{Dall'Agata:2010gj} is related to our $\phi$ by $\phi=2U$. Their gauge
coupling constants $g_{\Lambda}$ correspond to $2g_I$.}
\eq
W = e^{\phi/2}|{\cal Z} - i e^{2\psi-\phi}{\cal L}|\,, \label{W-Gianguido}
\feq
where ${\cal Z}=Q_IX^I-P^IF_I$ denotes the central charge, $Q_I$ and $P^I$ are the electric and
magnetic charges respectively, and ${\cal L}=2X^Ig_I$ is the actual superpotential defining the
scalar potential according to
\[
V = g^{\alpha\bar\beta}{\cal D}_{\alpha}{\cal L}{\cal D}_{\bar\beta}\bar{\cal L} - 3|{\cal L}|^2\,.
\]
Now, using $C_a=(Q_I,-P^I)^T$, one shows that the central charge obeys
\[
e^{\phi/2}{\cal Z}=(q^a-iH\Omega^{ab}q_b)C_a\,,
\]
which in our case boils down to $e^{\phi/2}{\cal Z}=q^aC_a$ since (as already pointed out above)
$C_a\Omega^{ab}q_b$ vanishes here. Furthermore, for axion-free solutions one has
$e^{\phi/2}g_IX^I=\mp ie^{\phi}k^aq_a/2$, and thus
\eq
e^{\phi/2}({\cal Z} - ie^{2\psi-\phi}{\cal L}) = q^aC_a \mp k^a q_a e^{2\psi}\,.
\feq
This coincides with \eqref{W-general} if we take the lower sign and ${S_a}^b={\delta_a}^b$,
or for the upper sign and ${S_a}^b=-{\delta_a}^b$ (plus an inversion $z\to-z$).
In conclusion, we saw that our fake superpotential \eqref{W-general}, which is
identical to Hamilton's characteristic function in a Hamilton-Jacobi formalism, matches in the
supersymmetric case (where $S=\pm\text{Id}$) the superpotential \eqref{W-Gianguido}
proposed in \cite{Dall'Agata:2010gj}.

For the $F=-X^1X^2X^3/X^0$ model considered in section \ref{-X1X2X3/X0-model}, things are more
subtle since we kept $x^0\neq 0$ such that the zero-axion condition is not satisfied. The analysis above
does therefore not apply. With only $Q_0$, $P^{1,2,3}$ nonvanishing, and the
truncation $F_0$, $X^{1,2,3}$ imaginary, \eqref{W-Gianguido} boils down to
\eq
W = \left|-\frac14 Q_0 e^{-\chi^0} + \sum_{\alpha=5}^7\left(\frac{C_{\alpha}}4 e^{-\chi^{\alpha}} -
g_{\alpha-4}e^{\chi^{\alpha}+2\psi}\right) + \frac{ig_0}2 e^{2\psi-\phi-\chi^0}\right|\,, \label{W-Gianguido'}
\feq
where we defined $q_0\equiv\exp\chi^0$. This expression has a piece containing the gauge
coupling constant $g_0$, which can never arise by considering the equations of motion alone
(as we did in this paper), since the scalar potential is independent of $g_0$. On the other hand,
in the Killing spinor equations, $g_0$ does appear. Ref.~\cite{Dall'Agata:2010gj} made use of
supersymmetry, which is the reason for the presence of the $g_0$-dependent term in \eqref{W-Gianguido'}.
If we ignored this term, we would have
\eq
W = \pm\left[\frac14 Q_0 e^{-\chi^0} + \sum_{\alpha=5}^7\left(-\frac{C_{\alpha}}4 e^{-\chi^{\alpha}} +
g_{\alpha-4}e^{\chi^{\alpha}+2\psi}\right)\right]\,, \label{W-Gianguido''}
\feq
which is, up to the $Q_0$-dependent piece and an irrelevant overall sign (that can be eliminated by
taking $z\to -z$) exactly \eqref{W-other-stu} with $\epsilon_{\alpha}=-1$. Our expression \eqref{W-other-stu}
is actually a reduced Hamilton's characteristic function in which the dynamics of $q_0$ was separated:
If we consider \eqref{W-Gianguido''} instead, we have (for the + sign)
\[
\dot\chi^0 = 4\frac{\partial W}{\partial\chi^0}\quad \Rightarrow \quad \dot q_0 = -Q_0\,.
\]
Comparing this with \eqref{q_0} in the limit $p\to 0$, one gets $Q_0=-D$. Notice that, in spite of $x^0\neq0$,
\eqref{W-other-stu} has again the form \eqref{W-general}, the reason for this being the decoupling of the
axion $x^0$ from the dynamics of the other fields. This, in turn, is a consequence of the independence
of the scalar potential of $g_0$.

\section{Conclusions and final remarks}
\label{conclusions}

One of our main results is the construction of a family of nonextremal black holes for the prepotential
$F=-2i(X^0X^1X^2X^3)^{1/2}$, which includes both the non-BPS solutions of \cite{Duff:1999gh},
and the supersymmetric black holes found in \cite{Cacciatori:2009iz}. We discussed some of their
physical properties, like entropy, temperature and several notions of mass, and showed that the
product of all horizon areas depends only on the charges and the asymptotic value of the cosmological
constant. We also provided a general recipe to construct non-BPS extremal solutions for an
arbitrary prepotential, as long as an axion-free condition holds. These follow from a set of first-order
conditions, and are related to the corresponding
supersymmetric black holes by a multiplication of the charge vector with a constant field rotation
matrix. We showed that the fake superpotential driving this first-order flow
coincides with Hamilton's characteristic function in a Hamilton-Jacobi formalism, and reduces
in the supersymmetric case to the superpotential proposed in \cite{Dall'Agata:2010gj}.

Some questions for future work are:
\begin{enumerate}
\item We saw in section \ref{nonextr-bh} that the mass \eqref{M-mAdS-eval} correctly vanishes
in the BPS case, but also for certain values of the parameters that do not correspond to any known
supersymmetric solution of ${\cal N}=2$ gauged supergravity. It would be interesting to see whether
they are BPS in extended supergravity theories, as conjectured in
\cite{Toldo:2012ec} for solutions of the $t^3$ model. In the ungauged case, this does indeed happen, and
was termed `camouflaged supersymmetry' in \cite{Bena:2011pi}.
\item Can we generalize the fake superpotential construction of section \ref{W-arbitr-prepot} to include
also axionic solutions? A hint how to do this may come from the true superpotential \eqref{W-Gianguido}
that works also in the axionic case.
\item Add rotation to the static nonextremal solutions constructed here. This is under investigation.

\end{enumerate}

\acknowledgments

This work was partially supported by INFN and MIUR-PRIN contract 2009-KHZKRX. The work of O.V.~is supported by an STFC studentship. D.K.~would like to thank Kiril Hristov for useful discussions.

\appendix

\section{Hesse potential for $F = -2i(Y^0Y^1Y^2Y^3)^{1/2}$ model}


We will consider the STU model characterised by the prepotential
\[
	F = -2i(Y^0 {Y^1} Y^2 Y^3)^{1/2} \;.
\]
The K\"ahler potential for this model can be written as
\[
	e^{-{\cal K}} = 8 Y^0 \bar{Y}^0 \text{Im}\left( i \sqrt{\frac{Z^2 Z^3}{Z^1}} \right)\text{Im}\left( i \sqrt{\frac{Z^1 Z^3}{Z^2}} \right)\text{Im}\left( i \sqrt{\frac{Z^1 Z^2}{Z^3}} \right) \;,
\]
where $Z^A = X^A / X^0$. As in the main text, we define $x^I := \text{Re}(Y^I)$ and $y_I := \text{Re}(F_I)$. The Hesse potential is related to the K\"ahler potential by $e^{-{\cal K}} = -2H$, and so in order to determine the Hesse potential we simply need to find an expression for the K\"ahler potential in terms of $x^I,y_I$.

By direct calculation one may show that
\begin{equation}
	\frac{y_3 i \sqrt{\frac{Z^2 Z^3}{Z^1 }} - x^2}{x^0 i \sqrt{\frac{Z^2 Z^3}{Z^1}} + y_1} = i\sqrt{\frac{\bar{Z}^1 \bar{Z}^2}{\bar{Z}^3}} \;,
	\label{eq:4}
\end{equation}
and since the calculation is identical if we interchange $X^2 \leftrightarrow X^3$ we also have
\begin{equation}
	\frac{y_2 i \sqrt{\frac{Z^2 Z^3}{Z^1 }} - x^3}{x^0 i \sqrt{\frac{Z^2 Z^3}{Z^1}} + y_1} = i\sqrt{\frac{\bar{Z}^1 \bar{Z}^3}{\bar{Z}^2}} \;,
	\label{eq:5}
\end{equation}
Again by direct calculation one may show that
\begin{equation}
	\frac{x^1 i \sqrt{\frac{Z^2 Z^3}{Z^1}} + y_0 }{x^0 i \sqrt{\frac{Z^2 Z^3}{Z^1}} + y_1} = \bar{Z}^1 \;.
	\label{eq:6}
\end{equation}
Notice that \eqref{eq:4} times \eqref{eq:5} yields minus \eqref{eq:6}, and so we obtain a quadratic
equation for $A := i(Z^2Z^3/Z^1)^{1/2}$, which is given by 
\[
	A^2 (\underbrace{x^0 x^1 + y_2 y_3}_a) + A (\underbrace{y_0 x^0 + y_1 x^1 - y_2 x^2 - y_3 x^3}_b) + \underbrace{y_0 y_1 + x^2 x^3}_c = 0 \;.
\]
We will only consider coordinate patches in which $4ac > b^2$, and so we can write the imaginary part of the solution as 
\begin{equation}
	\text{Im} \left( i \sqrt{\frac{Z^2 Z^3}{Z^1}} \right) = \pm \frac{\sqrt{4ac - b^2}}{2(x^0 x^1 + y_2 y_3)} \;. \label{eq:7}
\end{equation}
Given the symmetry of the model we can cycle through the coordinates $X^1\leftrightarrow X^2 \leftrightarrow X^3$, and so we also have 
\begin{align}
	\text{Im} \left( i \sqrt{\frac{Z^1 Z^3}{Z^2}} \right) &= \pm \frac{\sqrt{4ac - b^2}}{2(x^0 x^2 + y_1 y_3)} \;, \label{eq:8} \\
	\text{Im} \left( i \sqrt{\frac{Z^1 Z^2}{Z^3}} \right) &= \pm \frac{\sqrt{4ac - b^2}}{2(x^0 x^3 + y_1 y_2)} \;. \label{eq:9} 
\end{align}
A separate calculation allows us to express $X^0$ directly in terms of $x^I,y_I$,
\begin{equation}
	X^0 = \frac{x^0 i\sqrt{\frac{Z^2 Z^3}{Z^1}} + y_1}{\text{Im} \left( i\sqrt{\frac{Z^2 Z^3}{Z^1}}\right)} \;.
	           \label{eq:10}
\end{equation}
Substituting \eqref{eq:7}, \eqref{eq:8}, \eqref{eq:9} and \eqref{eq:10} into the expression for the K\"ahler
potential we find
\[
	e^{-{\cal K}} = 4 \sqrt{4ac - b^2} \;,
\]
where we have made appropriate sign choices such that the K\"ahler potential is strictly positive. The Hesse potential is therefore given by
\begin{eqnarray}
H(x,y) &=& -2\left(-(y_0x^0 - y_1x^1 - y_2 x^2 - y_3 x^3)^2 + 4y_1 x^1 y_2 x^2 + 4y_1 x^1 y_3 x^3\right.
                     \nonumber \\
	  & & \hspace{8em} \left.+ 4 y_2 x^2 y_3 x^3	+ 4 y_0 y_1 y_2 y_3 + 4 x^0 x^1 x^2 x^3\right)^{1/2} \,.
          	\label{eq:STUHesse}
\end{eqnarray}
Note that upon making the truncation 
	\begin{align*}
		x^0 &\to x^0 \;, 	& y_0 &\to y_0 \;, \\
		x^1 &\to x^1 \;, 	& y_1 &\to \tfrac13 y_1 \;, \\
		x^2 &\to x^1 \;, 	& y_2 &\to \tfrac13 y_1 \;, \\
		x^3 &\to x^1 \;, 	& y_3 &\to \tfrac13 y_1 \;,
	\end{align*}
one obtains the Hesse potential for the $t^3$ model found in \cite{Klemm:2012yg}.

We could alternatively have calculated this Hesse potential by making a symplectic rotation of the Hesse potential for $F = -X^1 X^2 X^3/X^0$, which was calculated in \cite{Mohaupt:2011aa}. The symplectic transformation that links these two descriptions of the STU model is precisely that which sends 
	\[
		(1,s,t,u,-stu,tu,su,st)^T \longmapsto (1,-tu,-su,-st,-stu,s,t,u)^T \;,
	\]
and is characterised by the symplectic matrix 
	\[
		{\cal S} = \left(\begin{array}{cccc|cccc} 
								+1 & 0 & 0 & 0 & 0 & 0 & 0 & 0  \\ 
								0 & 0 & 0 & 0 & 0 & -1 & 0 & 0  \\ 
								0 & 0 & 0 & 0 & 0 & 0 & -1 & 0  \\ 
								0 & 0 & 0 & 0 & 0 & 0 & 0 & -1  \\ \hline 
								0 & 0 & 0 & 0 & +1 & 0 & 0 & 0  \\ 
								0 & +1 & 0 & 0 & 0 & 0 & 0 & 0  \\ 
								0 & 0 & +1 & 0 & 0 & 0 & 0 & 0  \\ 
								0 & 0 & 0 & +1 & 0 & 0 & 0 & 0  
								\end{array} \right) \;.
	\]
The  coordinates $q^a = (x^I,y_I)^T$  transform as a vector under symplectic transformations,
	\[
		(x^0,x^1,x^2,x^3,y_0,y_1,y_2,y_3)^T \longmapsto (x^0,-y_1,-y_2,-y_3,y_0,x^1,x^2,x^3)^T \;.
	\]
The Hesse potential itself transforms as a scalar under symplectic transformations, $H(x,y) \mapsto H(x',y')$. 
One may check that applying this transformation to the Hesse potential for the $F = -X^1 X^2 X^3/X^0$ model given in \cite{Mohaupt:2011aa} results precisely in the expression \eqref{eq:STUHesse}.


\end{document}